\newcommand{\bra}[1]{%
{\left\langle #1\right\vert}}
\newcommand{\ket}[1]{%
{\left\vert #1\right\rangle}}
\newcommand{\braket}[2]{%
{\left\langle #1  \vert  #2\right\rangle}}

\newcommand{\set}[1]{%
{\{ #1 \}}}

\newcommand{\absq}[1]{%
{\vert #1 \vert^2}}

\newcommand{\bfm}[1]{%
{\mbox{\boldmath $#1$}}}

\documentclass{ws-ijmpb}
\begin{document}
\markboth{Masao Matsumoto}
{Restriction on types of coherent states 
due to gauge symmetry}
\title{Restriction on types of coherent states\\
due to gauge symmetry
\footnote
{The article is a slightly augmented and corrected version 
of Part V of the author's Doctoral Thesis 
accepted by Graduate School of Science and Engineering, 
Ritsumeikan University-BKC.}
}
\author{Masao Matsumoto
\footnote{Preferred
Mailing Address: 1-12-32 Kuzuha Asahi, Hirakara, 
Osaka 573-1111, Japan. 
For the sake of more effective communication, it is
preferable to use the address. }}
\address{Lecturer Anteroom, Shieikan, Ryukoku
University\\
Fukakusa Tsukamoto-cho 67, Fushimi-ku, Kyoto 612-8577, Japan
\\
matumoto@i.h.kyoto-u.ac.jp
\\
masao-matsumoto@m-pine-tree.sakura.ne.jp 
}
\maketitle

\begin{history}
\received{7 March 2013}
\revised{16 October 2013}
\accepted{14 January 2014}
Published 25 March 2014
\end{history}
%
\begin{abstract}
From the viewpoint of the SU(2) 
coherent states (CS) and their path integrals (PI) labeled by a full set of Euler
 angles $(\phi, \theta, \psi)$ which we developed in the previous paper,  
we study the relations between gauge symmetries of Lagrangians and 
allowed quantum states; 
we investigate permissible types of fiducial vectors (FV) in 
the full quantum dynamics in terms of SU(2) coherent states for typical Lagrangians. 
We propose a general framework for a Lagrangian having 
a certain gauge symmetry with respect to one of the Euler angles $\psi$. 
We find that for the case fiducial vectors are so restricted that they belong to 
the eigenstates of ${\hat S}_3$ 
or to the orbits of them under the action of the SU(2); 
and the strength of a fictitious monopole, 
which appears in the Lagrangian, is a multiple of $\frac12$. 
In this case Dirac strings are permitted. 
Our formulations and results deepen those of the preceding 
work by Stone that has piloted us; 
we illustrate the relation 
between the two methods. 
The reasoning here does not work for 
a Lagrangian without the gauge symmetry. 
This suggests a new possibility 
about monopole charge quantization. 
Besides analogies to field theory and entanglements 
in quantum information (QI) are briefly mentioned. \\


\end{abstract}
\renewcommand{\thefootnote}{\noindent \alph{footnote}}
\section{Introduction} 
\label{sec:intro}
Symmetry is one of the basic principles that 
penetrate all of physics: 
from classical to quantum physics
\cite{FeynLS1}${}^{\mbox{--}}$\cite{FonGhir}; 
from statistical or condensed matter
physics to particle physics
\cite{LLSP}${}^{\mbox{--}}$\cite{Aitch-GFT}; 
and from relativity to gauge field theory.
\cite{FeynLS1,Aitch-GFT} 
Thus we see a wide range of symmetries: from
external (space-time) to internal ones; 
and from discrete to continuous ones. 
This is partly because geometry is an indispensable 
element to describing physics. 
And a natural algebraic language to express 
geometrical symmetry is group theory.
\cite{FonGhir} 
\par
Classical mechanics is widely known 
to have close relations between symmetries and Lagrangians or Hamiltonians.
\cite{Lanczos,YourMand}
Since quantum mechanics and field theory are, 
in some respects, modeled and devised after classical mechanics, 
we see that not a few methods and notions, including 
 symmetry, in quantum mechanics 
and field theory resemble those in classical mechanics.
\footnote
{We all realize that the very methods of quantization also fall within such examples.} 
And thus symmetry plays a crucial role also in quantum physics. 
Or we can interpret 
that although ``the physical world is quantum mechanical'',\cite{Feyn} 
the quantum features are somewhat transmitted to the classical world; 
and through the latter we may try to grasp the former. 
Of course since we are not able to capture all Nature by
classical analogies, there are sometimes discrepancies 
between quantum symmetries and classical ones.
\cite{Jackiw,Bertl}
\par
Now, one of the typical mathematical tools that relate classical states 
with quantum ones is coherent states (CS).
\cite{KlaSk}
(In what follows we use each of the the acronyms as a plural as well as a singular.) It was originally devised by Schr{\"o}dinger
\cite{Schroed} as 
the states having classical ``particle'' nature. 
The state exhibits a wave packet whose center 
moves along with the classical trajectory 
with minimum uncertainty, thus it shows classical nature. 
The original CS which is called canonical CS is, in the light of quantum optics, 
generated by displacing, 
or driving, the vacuum, i.e. the zero photon state.
\cite{Glau-PR2}
Later CS have been developed in a wide variety of directions. 
Viewed from a general framework, we may take up the following 
three subjects among the evolutions. 
First, CS have been extended to wider classes. 
A systematic way to broaden CS is constructing CS 
in terms of unitary irreducible representations of Lie groups due to
Perelomov.\cite{Per-Book} 
In the approach, CS is defined by operating a unitary operator 
related to a physical system being considered on a ``fiducial vector 
(FV)'', which we denote $\ket{\Psi_0}$. 
From this point of view, for the canonical CS 
the unitary operator is a displacement operator 
and a FV is the ground state or 
vacuum. Similarly, the spin CS can be constructed by 
operating a rotation operator on a FV. 
The FV is conventionally taken as $\ket{s, s}$
or $\ket{s, - s}$: the highest or lowest eigenvectors of ${\hat S}_3$. 
We may perform the procedures to other Lie groups, 
which automatically produces CS for the corresponding Lie groups. 
Schur's lemma coming from  irreducibility always ensures the overcompleteness of CS.  
Second, since CS enjoy overcomplete relations, 
``coherent state path integrals (CSPI)'', 
i.e., path integrals (PI) via CS, have been developed.
\cite{KlaSk,Klaua}${}^{\mbox{--}}$\cite{IKG} 
Such CSPI have been pushing the method of PI forward strongly. 
And besides CSPI turned out to be closely related to geometric phases.
\cite{SW} 
In fact it is remarkable that geometric phases follow from the
topological terms of phase space PI or CSPI naturally.
\cite{KI}${}^{\mbox{--}}$\cite{KraC} 
Third, following the fruits of the above two developments, 
CS and CSPI with arbitrary FV have been explored.
\cite{NECSPI,spinPI3a} 
In the case CS are obtained by operating a unitary operator on an {\em arbitrary} FV: For the canonical CS a FV is arbitrary superpositions
of the Fock number states; and for the spin CS arbitrary superpositions 
of $\ket{s, m}$: a general eigenvector of ${\hat S}_3$. 
We found that, as mentioned in 
Ref. \refcite{spinPI3a}, the canonical CS evolving from 
arbitrary FV\cite{NECSPI} turned out to be an arbitrary superposition of displaced number states having no classical analogues. 
Similarly, we may regard spin CS evolving from a general FV other than the conventional one as 
quantum
 states without classical analogues. 
It is true that CS with the conventional FV are closest to classical states and have useful properties.
\cite{Per-Book} 
However, recent technologies enable us to prepare quantum states which have no classical analogues; 
the typical one is the squeezed states of light.
\cite{SQ} 
We certainly regard the evolutions plausible since experimental 
developments due to high technologies have often created opportunities 
to reconsider Nature. 
In this respect we may take CS evolving 
from a generic FV as the mathematical tools, or a sort of new language, 
for describing nonclassical quantum states. 
And thus what we have done in Refs. \refcite{NECSPI} 
and \refcite{spinPI3a} 
is constructing new quantum states and investigating
the dynamics: i.e., the time evolutions of the quantum 
states. 
We can interpret the attempts as extending both
CS due to Schr{\"o}dinger--Klauder--Glauber
--Perelomov and PI due to 
Dirac--Feynman--Klauder--Kuratsuji--Suzuki. 
\par
In the previous paper,\cite{spinPI3a} 
hereafter referred to as {\bf I}, 
we have developed a basic formulation of 
the SU(2), i.e. spin, CS based on arbitrary FV and 
of their PI. 
The CS and CSPI are labeled by a full set of 
three Euler angles ${\bf\Omega} \equiv (\phi, \theta, \psi)$. 
Since the present paper flows directly out of {\bf I}, 
we will look back the previous results concisely. 
In {\bf I} we found out that the Lagrangian in the action 
appeared in the PI expression were composed of two parts: 
The topological term related to geometric phases 
and the dynamical one originating from a Hamiltonian. 
And the former is again split into two 
parts: 
One is the monopole type part which is the generalization of that 
of Balachandran {\em et al.}
\cite{Bal-NP}${}^{\mbox{--}}$\cite{Aitch} 
and the other represents the effect of entanglements
between neighboring components of a FV. 
\footnote{The entanglements here 
do not concern those between states themselves that often employed in quantum computation (QC) and 
information (QI); instead, they relate to those between components of a given FV. However, see also the fourth point in 
future prospects in Sec. \ref{sec:discussion}. 
There we take up some similarities between 
our methods and the nature of entanglements 
in QI for reference frames 
in relativity.}
Such interweaving of components of
a FV appears in the dynamical term as well. 
The monopole is fictitious in that 
it does not represent a real physical monopole 
having a magnetic charge; instead it stems from the topological or geometric
phase terms. However, mathematical descriptions seem quite common to 
both real and fictitious monopoles. 
And we have confirmed the PI form by demonstrating 
from discrete PI to continuous ones. 
Moreover, it has been proved that the generic spin CSPI 
contract to the general canonical CSPI in the high spin limit.
\par
Now, we are going into a problem that 
we have given the advance notice in {\bf I}; 
see Sec. 6 in {\bf I}. 
At first sight it seems that we are free to choose FV; 
there are no restrictions on FV and we may take 
an arbitrary FV. 
However, when a Lagrangian varies at most a total derivative 
under a certain gauge transformation and possesses 
a sort of semiclassical symmetry, 
a full quantum state with an arbitrary FV 
does not always preserve the related symmetry. 
In the case, when a FV $\ket{\Psi_0}$ 
cannot be reached from $\ket{s, m}$ via ${\hat R}^{(s)}({\bf\Omega})$, 
we find some strange feature: 
Semiclassical orbits do not always 
represent exact quantal evolutions. 
It is Stone who first observed that 
there does exist one of the central problems 
at the point viewed from a general framework of CS with arbitrary FV. 
He raised the problem in a paper\cite{Stone} 
commenting on the precursory version of {\bf I}.
\cite{spinPIGP} 
Moreover, he went further enough to propose a criterion 
under which the CS capture full quantal evolutions. 
According to that, an arbitrary FV is not always 
realized and that there may be restrictions on FV 
so that quantum evolutions are consistent with the 
semiclassical ones which has the original symmetry. 
Actually, the FV have to be identical with $\ket{s, m}$ 
or on the orbits of $\ket{s, m}$ under the action of 
${\hat R}({\bf\Omega})$. 
And, as Stone precisely pointed out, the problem is deeply related 
to the charge quantization of monopoles. 
\par
In this article we consider the above problem posed in 
Ref. \refcite{Stone} from the general framework of spin CSPI developed in {\bf I}. 
This is precisely one of the topics that we promised 
to clarify at the end of {\bf I}. 
As mentioned earlier, we have demonstrated the process of 
going from the discrete PI to the continuous PI in {\bf I}. 
We have also showed that the spin CSPI contract to the canonical
CSPI. So the PI expressions in {\bf I} are quite all right and 
are not responsible for not bringing the restriction on FV. 
Then one might wonder from where the restriction comes. 
We will approach the riddle in the light of 
the ``gauge symmetry'' associated with the 
invariance of Lagrangian appeared in the spin CSPI 
in {\bf I}. 
So the present article is also concerned with ``symmetry''. 
It will be one of the early attempts 
that relate CS, CSPI and monopoles with symmetry, 
especially gauge symmetry.
\par
Let us enumerate what are new in the present paper 
concisely. It may be helpful to see how we will step forward from {\bf I}. 
First, we will see that the formulation of the spin CSPI in 
{\bf I}, when augmented with a subsidiary condition 
reflecting a gauge symmetry, actually brings 
the restriction on FV. 
Hence we will strengthen the validity 
of the results in {\bf I}. 
Second, during the course, we investigate the relation 
between semiclassical and full quantum-mechanical 
time evolutions quite thoroughly using concrete examples. 
And the method in Refs. \refcite{Bal-NP}--\refcite{Aitch} 
is extended to CS and CSPI with a more generic Lagrangian. 
Third, each of CS, CSPI, monopoles and gauge symmetry 
has been investigated so far. 
However, there seems to have been no attempts to 
combine all of them and indicate the relations between 
them. We will actually do it here in order to understand the 
restriction on FV in CS. 
Fourth, we find that the method of reasoning 
employed in gauge symmetry here does not apply to 
a Lagrangian without the symmetry. 
This suggests that there is a new possibility 
about fictitious monopole charge quantization. 
Fifth, we indicate that semiclassical or full quantum 
symmetries have close relations to other 
fundamental physical systems via Lie group 
formalisms. 
Sixth, we take up some similarities between our 
methods and other systems. 
They are field theoretic vacua, covariant quantization 
of photons and the nature of entanglements 
for reference frames in relativity.
\par
The plan of the paper is as follows. 
First, we look back the spin(SU(2))CSPI based on arbitrary FV in Sec. \ref{sec:spinCSPI}. 
Next, using the formulation of the spin CSPI, 
we discuss general properties of 
a Lagrangian in the light of gauge symmetries 
in Sec. \ref{sec:symLagFV}. 
Next, in Sec. \ref{sec:Lags} 
we demonstrate, using simple examples, the relations 
between types of FV and semiclassical as well as full quantal dynamics. 
We then look over several real examples of Lagrangians in order to 
see the relations between Hamiltonians, FV and gauge symmetries 
of the whole Lagrangians. 
Main results are presented as theorems and proved in 
Sec. \ref{sec:genTheorems}. 
Theorem \ref{th:typeFV} gives the central result concerning 
the restriction on FV in the full quantum picture. 
We find that gauge symmetries bring restrictions 
on FV and thus on the form of CS. 
We look into the situation much deeper by 
investigating the generator of the symmetry transformation in Theorem 
\ref{th:Gform}. 
Next, we revisit the gauge symmetries in the light of a new kind of 
isotropy subgroups due to Ref. \refcite{Stone} in 
Sec. \ref{sec:anotherViewSCevolvs}, 
and we see the correspondence between the approach 
and that in Secs. \ref{sec:symLagFV} and  \ref{sec:Lags}. 
Finally we summarize the results in Sec. \ref{sec:discussion}. 
There we also discuss several related topics that we view 
in a future prospect. 
\section{General SU(2) Coherent State Path Integrals}
\label{sec:spinCSPI}
Let us recall the results in {\bf I}. 
First, we define the spin(SU(2))CS, $\ket{\bf\Omega}$, 
evolving from an arbitrary FV $\ket{\Psi_0}$ as: 
\begin{equation}
\ket{\bf\Omega} \equiv \ket{\phi, \theta, \psi} 
= {\hat R}({\bf\Omega}) \ket{\Psi_0}
= \exp (- i \phi {\hat S_3}) \exp (- i \theta {\hat S_2})
 \exp (- i \psi{\hat S_3}) \ket{\Psi_0}. 
\label{eqn:CS1}          
\end{equation}
The FV $\ket{\Psi_0}$ is represented by: 
\begin{equation}
\ket{\Psi_0} = \sum_{m = - s}^{s} c_m \ket{m} 
\qquad {\rm with} \qquad
\sum_{m = - s}^{s} \absq{c_m} = 1.
\label{eqn:spinFV}
\end{equation}
Hereafter $\ket{m}$ stands for $\ket{s, m}$. 
From (\ref{eqn:CS1}) and (\ref{eqn:spinFV}) 
we obtain 
\begin{equation}
\ket{\bf\Omega} 
= \sum_{m=-s}^{s} c_m \ket{{\bf\Omega}, m} 
\qquad 
{\rm with} 
\qquad 
\ket{{\bf\Omega}, m} 
\equiv {\hat R}({\bf\Omega}) \ket{m}.
\label{eqn:CS3a}
\end{equation}
See ({\bf I}-19)
\footnote
{
Equation ({\bf I}-$\ast$) denotes Eq. ($\ast$) in {\bf I}.
} 
for the explicit form of $\ket{{\bf\Omega}, m}$ which
we do not need in the present paper. 
We called $\ket{{\bf\Omega}, m}$ 
the ``rotated spin number state'' in {\bf I} 
where we saw that it corresponded to the ``displaced number state''
\footnote
{
See, e.g., Ref. \refcite{DNS} and references therein.
}
in the general canonical CS.
\cite{NECSPI}
\par 
Then the quantum time evolution of a 
physical system with a Hamiltonian 
${\hat H} ({\hat S}_{+}, {\hat S}_{-}, {\hat S}_{3}; t)$ in terms of 
$\ket{\bf\Omega}$ is given by the propagator:
\begin{equation} 
K({\bf\Omega}_f, t_{f}; {\bf\Omega}_i, t_{i})
=\int 
\exp \{ (i / \hbar) S[{\bf\Omega}(t)] \} \, 
{\cal D} [{\bf\Omega}(t)],
\label{eqn:PI}
\end{equation}
where
\begin{equation}
S[{\bf\Omega}(t)]
\equiv 
 \int_{t_i}^{t_f} 
\Bigl[ \ \bra{\bf\Omega} i \hbar 
\frac{\partial}{\partial t} 
 \ket{\bf\Omega} 
 - H({\bf\Omega}, t) \ \Bigr] 
\, d t 
 \equiv \int_{t_i}^{t_f} 
 L({\bf\Omega}, {\bf \dot \Omega},t) \, d t, 
\label{eqn:action-pol}
\end{equation}
with 
\begin{equation}
H({\bf\Omega}, t) 
\equiv 
\bra{\bf\Omega} {\hat H} \ket{\bf\Omega}.
\label{eqn:H}
\end{equation}
The explicit form of the Lagrangian yields:
\begin{equation}
L({\bf \Omega}, {\dot {\bf\Omega}}, t) 
= \hbar \Bigl[ A_0(\set{c_m}) ({\dot \phi}\cos\theta + {\dot \psi}) 
+ A_3({\bf\Omega}, {\dot {\bf\Omega}}; \set{c_m}) \Bigr] 
- H({\bf\Omega}, t),
\label{eqn:Lagpol}
\end{equation}
where
\begin{equation}
A_3({\bf\Omega}, {\dot {\bf\Omega}}; \set{c_m}) 
\equiv 
- A_1(\psi; \set{c_m}) \, {\dot \phi} \sin\theta 
+ A_4(\psi; \set{c_m}) \, {\dot \theta}. 
\label{eqn:deffA3}
\end{equation}
The following expressions include 
what $A_0, A_1$ and $A_4$ mean: 
\begin{equation}
\left\{
\begin{array}{l}
A_0(\set{c_m}) 
=\sum_{m = - s}^{s} m \absq{c_m} 
\\
A_1(\psi; \set{c_m}) 
= (1 / 2) \sum_{m= - s + 1}^{s} f(s, m) 
[ c_m^{*} c_{m-1} \exp(i \psi) + c_m c_{m-1}^{*} 
\exp(- i \psi) ]
\\
A_2({\bf\Omega}; \set{c_m})
= (1 / 2) \sum_{m = - s + 1}^{s} f(s, m) 
\exp(i \phi) \{ (1 + \cos\theta) \exp(i \psi) c_m^{*} c_{m-1}
\\
\qquad \qquad 
 - (1 - \cos\theta) \exp(- i \psi) 
c_m c_{m-1}^{*} \} 
\label{eqn:defA}
\\
A_4(\psi; \set{c_m}) 
 \equiv 
[( 1 / (2 i)] \sum_{m=-s+1}^{s} f(s, m) 
[ c_m^{*} c_{m-1} \exp(i \psi) 
- c_m c_{m-1}^{*} \exp(- i \psi) ]. 
\\
f(s, m) 
= [(s + m)(s - m + 1)]^{1/2}. 
\end{array}
\right.
\end{equation}
\par
The term with the square brackets in 
the Lagrangian (\ref{eqn:Lagpol}):
\begin{equation} 
A_0(\set{c_m}) ({\dot \phi}\cos\theta + {\dot \psi}) 
+ A_3({\bf\Omega}, {\dot {\bf\Omega}}; \set{c_m}),
\label{eqn:topol-term0}
\end{equation}
stemming from $\bra{\bf\Omega}(\partial / \partial t) 
\ket{\bf\Omega}$, 
may be called the ``topological term'' 
that is related to the geometric phases. 
%
\par
For operators ${\bfm S} \equiv ({\hat S}_1, {\hat S}_2, {\hat S}_3)$ 
that constitute $\hat H$, we
have 
\begin{equation}
\left\{
\begin{array}{l}
\bra{\bf\Omega} 
{\hat S}_{3} 
\ket{\bf\Omega} 
= A_0(\set{c_m}) \cos\theta 
- A_1(\psi; \set{c_m}) \sin\theta 
 \\
\bra{\bf\Omega} 
{\hat S}_{+} 
\ket{\bf\Omega} 
= A_0(\set{c_m}) \sin\theta \exp(i \phi) 
+ A_2({\bf\Omega}; \set{c_m}) 
= \bra{\bf\Omega} 
{\hat S}_{-} 
\ket{\bf\Omega}^{*},
\end{array}
\right.
\label{eqn:mat-pol}
\end{equation}
where 
$A_i\, (i = 0, 1, 2)$ and $f(s, m)$ 
have already been given in (\ref{eqn:defA}) and 
$
{\hat S}_{\pm} = {\hat S}_1 \pm i {\hat S}_2 
$. 
\par
Variational equations associated with the 
Lagrangian (\ref{eqn:Lagpol}) are: 
\begin{equation}
\left
\{
\begin{array}{l}
\hbar 
\{ 
[A_0(\set{c_m}) \sin\theta 
+ A_1(\psi; \set{c_m}) \cos\theta 
] {\dot \phi} 
+ A_1(\psi; \set{c_m}) {\dot \psi}
\} 
= - ( \partial H / \partial \theta )
\\
\hbar 
\{
[
A_0(\set{c_m}) \sin\theta 
+ A_1(\psi; \set{c_m}) \cos\theta 
] {\dot \theta} 
- [A_4(\psi; \set{c_m}) \sin\theta] {\dot \psi} 
\} 
= \partial H / \partial \phi
\\
\hbar 
\{
[A_4(\psi; \set{c_m}) \sin\theta] {\dot \phi} 
+ A_1(\psi; \set{c_m}) {\dot \theta} 
\} 
= {\partial H / \partial \psi},
\label{eqn:vareq-pol} 
\end{array}
\right.
\end{equation}
\section{Gauge Symmetry of Lagrangian}
\label{sec:symLagFV}
Following the results in Sec. \ref{sec:spinCSPI}, 
we now consider the relations 
between semiclassical time evolutions 
of $\ket{\bf\Omega}$ and the properties 
of symmetries that Lagrangians possess.
\par
We see from (\ref{eqn:Lagpol})--(\ref{eqn:defA}) 
that the $\psi$-variable does not take effect 
in $A_i, \ (i = 1, 2, 4)$ provided no neighboring 
$\set{c_m}$ exists for any $c_m$. 
Then $A_3$-terms vanishes and the topological term 
(\ref{eqn:topol-term0}) takes the form:
\begin{equation}
A_0(\set{c_m}) ({\dot \phi}\cos\theta + {\dot \psi}). 
 \label{eqn:topTermSym} 
\end{equation}
The form of (\ref{eqn:topTermSym}) is a generalization of that 
in Ref. \refcite{Bal-CTQS}. 
Taking $s = \frac12, c_{1 / 2} = 1$ in (\ref{eqn:topol-term0}) 
yields the latter form. 
For such a FV, if we assume that $\hat H$ is linear in 
${\bfm S}$, 
we see from (\ref{eqn:mat-pol}) that $H({\bf\Omega})$ does not depend 
on $\psi$. 
Consequently the form of the variational equations (\ref{eqn:vareq-pol}) 
becomes the same as that for the usual spin CSPI evolving from 
$\ket{\Psi_0} = \ket{s}$, or $\ket{-s}$ with $\pm s$ replaced with $A_0$. 
And thus we can choose any $\psi$ as far as semiclassical dynamics is concerned. 
In what follows we will investigate a little deeper such FV and $\hat H$ 
as above-mentioned that yield (\ref{eqn:topTermSym}) and leave semiclassical dynamics invariant.
\par
From the viewpoint of the symmetry of Lagrangian, 
we grasp the situation as follows. 
To begin with, in the present case we have 
\begin{equation}
L({\bf\Omega}, {\bf \dot \Omega},t) 
= \hbar A_0 (\set{c_m}) 
({\dot \phi}\cos\theta + {\dot \psi}) 
- H({\bf\Omega}, t) .
\label{eqn:weak-sym-Lag}
\end{equation}
Under the ``gauge $\psi$-transformation'':
\begin{equation}
{\hat R}({\bf\Omega})
\longrightarrow {\hat R}({\bf\Omega}) \cdot 
\exp(- i {\hat S}_3 \psi') 
= {\hat R}({\bf\Omega}'), 
\qquad \bigl( {\bf\Omega}' 
\equiv (\phi, \theta, \psi + \psi') \bigr), 
\label{eqn:transfRS3}
\end{equation}
which moves the $\psi$-Euler angle, 
a ket vector $\ket{\bf\Omega}$ changes as: 
\begin{equation}
\ket{\bf\Omega} 
\equiv {\hat R}({\bf\Omega}) \ket{\Psi_0} 
\longrightarrow
{\hat R}({\bf\Omega}) \exp(- i {\hat S}_3 \psi') \ket{\Psi_0}
\equiv \ket{\bf\Omega'}.
\label{eqn:transf-ketV}
\end{equation}
And besides we assume 
\begin{equation}
H({\bf\Omega}) = H({\bf\Omega}').
\label{eqn:HH'}
\end{equation}
Then the Lagrangian (\ref{eqn:weak-sym-Lag}) behaves as: 
\begin{equation}
L \rightarrow L + \hbar A_0 {\dot \psi'}.
\label{eqn:transf-Lag}
\end{equation}
Hence we see that the Lagrangian (\ref{eqn:weak-sym-Lag}) 
has what is called ``weak invariance'' in 
Refs. \refcite{Bal-NP} and \refcite{Bal-CTQS}: 
The Lagrangian varies only by a total derivative term under 
a certain transformation.
\cite{Bal-NP}${}^{\mbox{--}}$\cite{Aitch} 
We have extended their method to our spin CS and CSPI 
with a more generic Lagrangian in (\ref{eqn:weak-sym-Lag}). 
The concrete examples of Lagrangians will be given in the next section. 
Let us call the symmetry the ``weak gauge symmetry'' 
or the ``weak gauge $\psi$-symmetry'' here 
so as to stress the effect of the gauge $\psi$-transformation
(\ref{eqn:transfRS3}). 
We state them once more in the following definition:
\begin{definition}
If a Lagrangian is transformed according to (\ref{eqn:transf-Lag}) 
under (\ref{eqn:transfRS3}), we say that 
the Lagrangian possesses the weak gauge symmetry or 
the weak gauge $\psi$-symmetry.
\end{definition}
\par
So far we have concentrated on the semiclassical time evolutions. 
We see that a FV with certain conditions meet the symmetry 
of semiclassical motions. 
However, if we consider the full quantum dynamics 
conformable to the $\psi$-invariance, 
more stringent conditions are required of FV. 
Moreover, we will find that the
conditions on FV, i.e. those on $\set{c_m}$, 
for the symmetry of Hamiltonians or of the whole 
Lagrangians are changed 
when we take up Hamiltonians that are quadratic or higher 
in ${\hat S}_\pm$. 
We will look into the 
situations more deeply using real sample Lagrangians 
in Sec. \ref{sec:Lags}.
\section{Sample Lagrangians}
\label{sec:Lags}
In this section we will discuss the relation 
between FV, Hamiltonians and weak symmetry of Lagrangians above 
mentioned by demonstrating several concrete examples. 
The problem is closely related to semiclassical versus 
full quantum time evolutions first pointed by Stone.
\cite{Stone} 
\par
First, we investigate a few simple examples 
to see how FV relate to symmetries 
and what semiclassical and full quantum time evolutions look like 
in Sec. \ref{sec:ex-simpIe-llusts}. 
This may help one to grasp various wider examples 
collected in Table \ref{tab:exLagsSym} in the
following Sec. \ref{sec:ex-Lags}. 
The examples in Sec. \ref{sec:ex-simpIe-llusts} 
are also included in Table \ref{tab:exLagsSym}.
\subsection{Simple illustrations}
\label{sec:ex-simpIe-llusts}
In this subsection we illustrate, using simple examples 
of a Hamiltonian and FV, the relations between 
semiclassical paths and exact quantal time evolutions. 
\par
Before going into the examples, 
let us recollect the general theory of CS.
\cite{Per-Book} 
Then we find that CS are determined 
in connection with FV. 
Let $\cal G$ be a Lie group of our concern. 
Consider the case in which there exists a subgroup of 
$\cal G$, say $\cal H$, that leaves a FV, 
$\ket{\Psi_0}$, invariant: 
\begin{equation}
{\cal H} \ket{\Psi_0} = \exp(i \alpha) \ket{\Psi_0}
\ (\alpha {\rm : phase}).
\label{isotropySG}
\end{equation}
Such a subgroup $\cal H$ is called the ``isotropy subgroup'' 
or the ``stabilizer''. 
Then CS is actually defined on the coset space 
${\cal G} / {\cal H}$. 
\footnote
{
Usually $G$ and $H$ are frequently employed 
in place of $\cal G$ and $\cal H$. 
In order that one does not confuse them with the Hamiltonians 
${\hat H}$, $H$ and the generator $\hat G$ 
in the following sections we use calligraphic $\cal G$ 
and $\cal H$ to express general groups here. 
} 
%
If there is no isotropy subgroups, we may regard 
${\cal H} = \set{\bf 1}$. 
Then ${\cal G} / {\cal H}$ is ${\cal G}$ itself. 
Hence we see that an isotropy subgroup depends upon 
the way in which we choose a FV, $\ket{\Psi_0}$. 
\par
Consider the present SU(2) CS case: 
${\cal G} = SU(2)$. 
If a FV is $\ket{\Psi_0} = \ket{m}$, 
then ${\cal H} = U(1)$ 
and the spin CS is actually determined on the coset space 
${\cal G} / {\cal H} = SU(2) / U(1)$ which 
corresponds to the Bloch sphere $S^2$, 
and thus $\ket{\bf\Omega} \rightarrow \ket{\theta, \phi}$. 
This includes what we always do in constructing the conventional 
spin CS with $\ket{\Psi_0}= \ket{s}$ or $\ket{- s}$, 
and we may call the FV ``standard''. 
The corresponding CS fall within what is called the ``informative'' 
CS in Ref. \refcite{Stone}. 
We see that CS is invariant under the transformation 
${\cal H} = \set{\exp(- i \psi {\hat S}_3)}$.
\footnote
{
We take ``$- i \psi$'' after our Euler angle convention.
} 
The case is illustrated in Sec. \ref{sec:sampleFV1} below.
Next, if we take a FV other than 
$\ket{\Psi_0} = \ket{m}$, 
then $\cal H$ is not $U(1)$ no more but $\set{\bf 1}$. 
Hence ${\cal G} / {\cal H} = {\cal G} = SU(2) \simeq S^3$ is specified 
by a {\em full} set of three Euler angles: 
${\bf\Omega} = (\phi, \theta, \psi)$. 
For this time $\exp(- i \psi {\hat S}_3) \ket{\Psi_0} 
\ne \exp(i \, \alpha) \ket{\Psi_0}$ ($\alpha$: {\rm phase}). 
However, there are cases in which the ``little group'' 
\cite{Aitch-GFT} that leaves semiclassical states 
 invariant exists. The group turns out to be $\set{\exp(- i \psi' {\hat G})}$
\footnote
{We take a minus sign for the convenience of later arguments.} 
where $\hat G$ is the generator of the weak symmetry 
transformation (\ref{eqn:transfRS3}). 
The explicit form will be given later in Sec. \ref{sec:explicit-G}. 
See Sec. \ref{sec:sample-nonStandardFV1} 
for a sample FV. 
The case has something to do with 
another type of isotropy subgroups proposed by Stone 
\cite{Stone}, and we will investigate the case 
from a slightly different viewpoint in Sec.
\ref{sec:anotherViewSCevolvs}. 
The case without even the weak symmetry 
is treated in Sec. \ref{sec:sample-nonStandardFV2}. 
\par
Now, we will treat concrete examples below.
The examples, which are discussed by Stone,
\cite{Stone}
are simplified versions of 
Ref. \refcite{spinPIGP}. 
Take a spin in a constant magnetic field 
${\bfm B} = (0, 0, B)$. 
The Hamiltonian is:
\begin{equation}
{\hat H} = - \mu B {\hat S}_3.
\label{eqn:HS3}
\end{equation}
In what follows we will illustrate, 
taking three typical types of FV, 
the relations between Hamiltonians, FV, 
the $\psi$-symmetry of Lagrangians. 
We also demonstrate how semiclassical 
motions concern the full quantum ones 
for the FV.
\subsubsection{Standard FV}
\label{sec:sampleFV1}
First, let a FV $\ket{\Psi_0} = \ket{m}$. 
It is rather a standard case where ${\cal H} =U(1)$. 
We have $A_0 = m$ and the $A_3$-term is absent, 
and thus the topological term has the weak 
$\psi$-symmetry as we saw it in Sec. \ref{sec:symLagFV}. 
Besides since the $A_1$- and $A_2$-term vanish,
\begin{equation}
H({\bf\Omega}) = - m\, \mu B \, \cos\theta
\label{eqn:H-FVm}
\end{equation}
contains no $\psi$-variables and 
is symmetric under (\ref{eqn:transfRS3}).
Consequently, the Lagrangian:
\begin{equation}
L = m \, ({\dot \phi} \cos\theta + {\dot \psi}) 
+ m\, \mu B \, \cos\theta
\label{eqn:FVmLag}
\end{equation}
possesses the weak gauge $\psi$-symmetry 
(\ref{eqn:transf-Lag}). 
The above situation is summarized in the column (i) 
in Table \ref{tab:exLagsSym}; 
see the following Sec. \ref{sec:ex-Lags}. 
\par
The variation equations (\ref{eqn:vareq-pol}) are:
\begin{equation}
{\dot \phi} = - (\mu B / \hbar), 
\qquad 
{\dot \theta} = 0.
\label{eqn:vareq-HS3}
\end{equation}
The third equation in (\ref{eqn:vareq-pol}), 
describing the behavior of $\dot \psi$, 
is automatically satisfied. 
It means that any $\psi$ works well 
as far as the semiclassical motions are concerned. 
From (\ref{eqn:vareq-HS3}) we obtain, 
setting the initial state as ${\bf\Omega}(t = 0) 
\equiv {\bf\Omega}_0
= (0, \theta_0, \psi_0)$:
\begin{equation}
\phi = - (\mu B / \hbar) t, 
\qquad 
\theta = \theta_0, 
\qquad 
\psi = \psi(t) \quad (\psi(t) {\rm: arbitrary}).
\label{eqn:vareq-HS3-sols}
\end{equation}
Therefore the semiclassical motion is given by:
\begin{equation}
\ket{{\bf\Omega}_{\rm SC}(t)} 
= \exp \{- i [(\mu B) / \hbar] 
{\hat S}_3 t \} 
\cdot \exp(- i \theta_0 {\hat S}_2)
\cdot \exp[- i \psi(t)] 
\ket{m}.
\label{eqn:vareq-HS3-sols-tot}
\end{equation}
\par
On the other hand, we see that the full quantal time evolution 
is also described in terms of spin CS with the Euler angle 
${\bf\Omega}_{\rm FQ}(t)$: 
\begin{eqnarray}
\ket{{\bf\Omega}_{\rm FQ}(t)} 
& = & 
\exp \{- i [(\mu B) / \hbar] \, 
{\hat S}_3 \, t \} \ket{{\bf\Omega}(0)}
\nonumber\\
& = & \exp \{- i [(\mu B) / \hbar] \, 
{\hat S}_3 \, t \} 
\cdot \exp(- i \theta_0 {\hat S}_2) 
\cdot \exp(- i \psi_0) 
\ket{m}.
\label{eqn:time-evol-HS3-m}
\end{eqnarray} 
For the present FV, i.e., $\ket{m}$, 
each factor, $\exp[- i \psi(t) {\hat S}_3]$ in 
(\ref{eqn:vareq-HS3-sols-tot}) and $\exp(- i \psi_0 {\hat S}_3)$ in (\ref{eqn:time-evol-HS3-m}), 
when acting on $\ket{m}$, yields a trivial phase factor respectively. 
And thus two states (\ref{eqn:vareq-HS3-sols-tot}) and 
(\ref{eqn:time-evol-HS3-m}) belong to the same ray 
living on $SU(2) / U(1) \simeq S^2$. 
Consequently, both semiclassical and
genuine quantum time evolutions coincide. 
We may interpret that the gauge $\psi$-symmetry 
is preserved also in the full quantum dynamics. 
\par
Next, let us see the relation between the full quantum propagator 
and the semiclassical one. 
For an arbitrary final state $\ket{{\bf\Omega}_f}$ 
at $t = t_f$ they are given by 
$\braket{{\bf\Omega}_f}{{\bf\Omega}_{\rm FQ}(t)}$ 
and $\braket{{\bf\Omega}_f}{{\bf\Omega}_{\rm SC}(t)}$ respectively.
Then it is clear that from (\ref{eqn:vareq-HS3-sols-tot}) and 
(\ref{eqn:time-evol-HS3-m}) the relation between two propagators are:
\begin{equation}
\braket{{\bf\Omega}_f}{{\bf\Omega}_{\rm FQ}(t)}
= \braket{{\bf\Omega}_f}{{\bf\Omega}_{\rm SC}(t)} 
\cdot \exp\{ i \, m \, [ \psi(t) - \psi_0] \}.
\label{eqn:diff2props-FVm}
\end{equation}
The semiclassical propagator 
obeys the full quantal one up to the $\psi$-gauge dependence. 
What we see here is one of the concrete examples of 
the dynamics for informative CS.\cite{Stone} 
Notice that for a round trip in which 
the final $\psi$ differs from $\psi_0$ by $2 \pi$ or 
$4 \pi$ two propagators fully agree since $m$ is an integer 
or a half integer. 
\subsubsection{Nonstandard FV1}
\label{sec:sample-nonStandardFV1}
Second, we put FV as: 
$\ket{\Psi_0} = (\sqrt{2/3}, 0, \sqrt{1/3})^T \notin
{\hat R}({\bf\Omega}) \ket{m} \ (m = 1, 0, -1)$.
\footnote
{
One may confirm that 
$\ket{\Psi_0} \notin {\hat R}({\bf\Omega}) \ket{m}$ 
holds actually with the aid of the explicit form of 
${\hat R}(\bf\Omega)$. 
See, e.g., Messiah \cite{Messiah} and references cited in {\bf I}. 
} 
As the previous FV only $A_0$-term survives among $A_i\, (i = 0, \cdots,
4)$ since the terms involving $c_m^* c_{m - 1}$ and its complex
conjugate vanish. 
The Lagrangian reads:
\begin{equation}
L = \frac13 ({\dot \phi} \cos\theta + {\dot \psi}) 
+ \frac13 \mu B \cos\theta.
\label{eqn:Lag-nontrivialFV1}
\end{equation}
It behaves like (\ref{eqn:transf-Lag}), showing 
the weak gauge $\psi$-symmetry. 
See the column (ii) in Table \ref{tab:exLagsSym} 
in the following Sec. \ref{sec:ex-Lags}. 
The variation equations are the same as (\ref{eqn:vareq-HS3}) 
and the solution gives:
\begin{equation}
\ket{{\bf\Omega}_{\rm SC}(t)} = \exp \{- [(\mu B) / \hbar] 
\, {\hat S}_3 \, t \} 
\cdot \exp(- i \theta_0 {\hat S}_2) 
\cdot 
\begin{pmatrix} \exp[- i \psi(t) ]\sqrt{2 / 3} \\ 0 \\ 
\exp[ i \psi(t) ] \sqrt{1 / 3} 
\end{pmatrix}.
\label{eqn:vareq-HS3-sols-tot-ntFV}
\end{equation}
Since the full quantum time evolution operator 
is clearly the same as that in 
(\ref{eqn:time-evol-HS3-m}), 
the corresponding state is given by:
\begin{equation}
\ket{{\bf\Omega}_{\rm FQ}(t)} = \exp \{- [(\mu B) / \hbar] 
\, {\hat S}_3 \, t \} 
\cdot \exp(- i \theta_0 {\hat S}_2) 
\cdot 
\begin{pmatrix} \exp(- i \psi_0 ) \sqrt{2 / 3} \\ 0 \\ 
\exp(i \psi_0) \sqrt{1 / 3} 
\end{pmatrix}.
\label{eqn:time-evol-HS3-ntFV1}
\end{equation}
We realize that this time each factor, 
$\exp[- i \psi(t) {\hat S}_3 ]$ 
in (\ref{eqn:vareq-HS3-sols-tot-ntFV}) 
and $\exp(- i \psi_0 {\hat S}_3)$ in (\ref{eqn:time-evol-HS3-ntFV1}), 
when acting on the present
 FV, yields a {\em nontrivial}
 phase factor that changes the
 quantum states basically. 
If we have $\psi(t) \equiv \psi_0$ for all $t$, two evolutions 
coincide. However, there seems no {\em a priori} reason 
to set $\psi(t) \equiv \psi_0$; 
for we know that the Lagrangian 
(\ref{eqn:Lagpol}) has the gauge $\psi$-symmetry. 
And thus we conclude that the semiclassical time evolution does not agree with 
the genuine quantum time evolution for this {\em nontrivial} FV as pointed out by Stone.
\cite{Stone} 
Notice that the relation between full quantum propagators and the
semiclassical one is no longer simple as 
(\ref{eqn:diff2props-FVm}), but we have instead:
\begin{equation}
\braket{{\bf\Omega}_f}{{\bf\Omega}_{\rm FQ}(t)} 
= \bra{{\bf\Omega}_f} 
{\hat R}({\bf\Omega}_{\rm SC}) 
\vert {\tilde \Psi}_0 \rangle
\quad {\rm with} \quad 
\vert {\tilde \Psi}_0 \rangle
\equiv \sum_{m = - 1}^{1} c_m \exp[i\, m ( \psi(t) - \psi_0)] \ket{m}.
\label{eqn:ediff2props-nonTrivialFV1}
\end{equation}
\subsubsection{Nonstandard FV2}
\label{sec:sample-nonStandardFV2}
Third, we consider the FV: 
$\ket{\Psi_0} = (\sqrt{1 / 2}, \sqrt{1 / 6}, 
\sqrt{1 / 3})^T \notin
{\hat R}({\bf\Omega}) \ket{m} \ (m = 1, 0, -1)$. 
See the column (v) in Table \ref{tab:exLagsSym} 
in the following Sec. \ref{sec:ex-Lags} 
and one will find the both $A_3$-term and $H$ 
have different properties from two above-mentioned 
FV cases, thus yielding 
the Lagrangian without the $\psi$-symmetry. 
We have the Lagrangian:
\begin{equation}
L = \frac16 \hbar \{ 
 ({\dot \phi} \cos\theta + {\dot \psi}) 
+ {\tilde c} \, [ {\dot \theta} \sin\psi - {\dot \phi}
\sin\theta\cos\psi ] \} 
 - \frac16 \mu B [ \cos\theta - {\tilde c} \sin\theta] 
\cos\psi, 
\label{eqn:Lag-nontrivialFV3}
\end{equation} 
where ${\tilde c} \equiv 2 + \sqrt{6}$. 
The variation equations (\ref{eqn:vareq-pol}) give:
\begin{equation}
\left
\{
\begin{array}{l}
\hbar 
\{
[\sin\theta + {\tilde c}\cos\theta \cos\psi] {\dot \phi}
+ {\tilde c}{\dot \psi} \cos\psi 
\}
= - \mu B (\sin\theta + {\tilde c}\cos\theta \cos\psi),
\\
\hbar 
\{
[\sin\theta + {\tilde c}\cos\theta \cos\psi] {\dot \theta}
 - ({\tilde c}\cos\theta \sin\psi) {\dot \psi} 
\} = 0,
\\
\hbar 
({\dot \phi} \sin\theta \sin\psi 
 + {\dot \theta} \cos\psi) 
= - \mu B \sin\theta \sin\psi. 
\label{eqn:vareq-nonStandardFV2} 
\end{array}
\right.
\end{equation}
One can verify that (\ref{eqn:vareq-nonStandardFV2}) 
yields a solution: 
\begin{equation} 
\phi = - (\mu B / \hbar) t, 
\qquad 
\theta = \theta_0,
\qquad 
\psi = \psi_0 \quad 
(\psi_0 {\rm : const.}),
\label{eqn:vareq-HS3-sols-nonStandardFV2}
\end{equation} 
which provides the same time evolution 
of the state as that for the exact quantum dynamics. 
For we know that the time development operator acting on a FV 
for the latter is the same as that in (\ref{eqn:time-evol-HS3-m}) 
or (\ref{eqn:time-evol-HS3-ntFV1}).
\subsection{Various examples of Lagrangians}
\label{sec:ex-Lags}
When a Hamiltonian $\hat H$ includes terms that are quadratic or higher
in ${\hat S}_\pm$, 
$\psi$-dependence of $H$ will be changed from that 
for $\hat H$ linear in ${\hat S}_i (i = \pm, 3)$ even for the same FV. 
And thus the gauge $\psi$-symmetry of the whole Lagrangian 
will be also modified. 
We have to be careful that 
{\em generally in CSPI 
the combination of a Hamiltonian 
and a FV together determines 
whether the corresponding Lagrangian possesses 
the weak gauge $\psi$-symmetry or not.} 
This is a crucial difference from Refs. 
\refcite{Bal-NP}--\refcite{Aitch}. 
In the point two cases are completely different. 
\par
In the present subsection we take up two typical Hamiltonians that 
are composed of $\bfm S$. 
They are the NMR and NQR types of Hamiltonians 
given below:
\cite{Slichter,Tycko} 
\begin{equation}
{\hat H}_{\rm NMR} 
= - \mu {\bfm B} {\bf \cdot} {\bfm S}
\qquad ({\rm NMR \ \ type})
\label{eqn:H-NMR}
\end{equation}
and
\begin{equation}
{\hat H}_{\rm NQR} 
= \omega_Q ({\bfm B}{\bf \cdot}{\bfm S})^2 
\qquad ({\rm MQR \ \ type}).
\label{eqn:H-NQR}
\end{equation}
We can obtain $H$ in (\ref{eqn:H}) as 
$H_{\rm NMR} \equiv \bra{\bf\Omega} 
{\hat H}_{\rm NMR} \ket{\bf\Omega}$. 
We may have the explicit form with the aid of 
(\ref{eqn:mat-pol}). 
Similarly $H_{\rm NQR}$ is evaluated with the aid of 
({\bf I}-A.8). For the present purpose, however, we do not 
have to know the explicit forms of $H_{\rm NMR}$ and 
$H_{\rm NQR}$. 
All that we need to grasp is the following: 
$H_{\rm NMR}$ is independent of $\psi$ and hence invariant under 
(\ref{eqn:transfRS3}) if and only if no nearest neighboring 
$\set{c_m}$ exists for any $c_m$. 
And the condition for $H_{\rm NQR}$ to be invariant 
under (\ref{eqn:transfRS3}) is that 
both no nearest neighboring 
$\set{c_m}$ 
and no next nearest neighboring $\set{c_m}$ in a given FV
exist for any $c_m$. 
\par
We demonstrate various examples of combinations 
of Hamiltonians and FV, 
including those illustrated in Sec. \ref{sec:ex-simpIe-llusts}, 
 in Table \ref{tab:exLagsSym} below, 
and we indicate the semiclassical weak gauge $\psi$-symmetry of their topological term, $H$ and the whole Lagrangian. As we put in Sec. \ref{sec:ex-simpIe-llusts}, 
all the examples that meet the semiclassical symmetry 
do not always preserve the symmetry in full quantum dynamics. We will proceed to the problem in the next section.
\par
In Table \ref{tab:exLagsSym} notations like 
$(\sqrt{2 / 3}, 0, \sqrt{1 / 3})^T 
\equiv (\frac23)^{1/2} \ket{1} 
+ (\frac13)^{1/2} \ket{- 1}$, 
for instance, are used. 
And Hamiltonians (\ref{eqn:H-NMR}) and 
(\ref{eqn:H-NQR}) are referred to as NMR 
and NQR respectively. 
Notice that in case (i) the spin actually meets $s \ge 1$ 
for NQR type Hamiltonians.
\cite{Slichter}
\begin{table}[tphb] 
\tbl{Examples of Lagrangians and weak gauge symmetry}
{\begin{tabular}{@{}cccccccc@{}} \Hline 
\\[-1.8ex] 
 & (i) & (ii) & (iii) & (iv) & (v) & (vi) & (vii)\\
\hline \\[-1.8ex]
\\ 
 spin & arbitrary & $1$ & 1 & 1 & 
 1 & $3 / 2 $ & $3 / 2$ \\
 & & & & & & & \\ 
$\hat H$ & NMR & NMR & NQR 
& NMR & NMR & NMR & NQR \\
{} & NQR & {} & {} & {} & {}& {} \\
\\ 
FV $\ket{\Psi_0}$ & $ \ket{m}$
& $\begin{pmatrix} \sqrt{2 / 3} \\ 0 \\ 
\sqrt{1 / 3} 
\end{pmatrix}$ & 
$\begin{pmatrix} \sqrt{2 / 3} \\ 0 \\ 
\sqrt{1 / 3} 
\end{pmatrix}$ 
& $\begin{pmatrix} \sqrt{1 / 3} \\ 
\sqrt{1 / 3} \\ 
\sqrt{1 / 3} 
\end{pmatrix}$ & 
$\begin{pmatrix} \sqrt{1 / 2} \\ 
\sqrt{1 / 6} \\ 
\sqrt{1 / 3} 
\end{pmatrix}$ 
& $\begin{pmatrix} \sqrt{2 / 3} \\ 0 \\ 0 \\
\sqrt{1 / 3} \end{pmatrix}$ 
& $\begin{pmatrix} \sqrt{2 / 3} \\ 0 \\ 0 \\
\sqrt{1 / 3} \end{pmatrix}$ \\
\\
$A_0$ & $m$ & $1 / 3$ & $1 / 3$ & $0$ & 
 $1 / 6$ & $1 / 2$ & $1 / 2$ \\
\\
 $A_3$-term & absent & absent & absent 
& present & present & absent & absent \\
\\ 
 weak symmetry & & & & & & & \\
of the & Yes & Yes & Yes & No & No & Yes & Yes\\
topological term 
\\
 \\
 symmetry & & & & & & & \\
of $H$ & Yes & Yes & No
& No & No 
& Yes & Yes \\
\\
total weak & & & & & & & \\
 symmetry & Yes & Yes & No & No & No & Yes & Yes\\
of $L$ & & & & & & & \\
\\
\Hline\\[-1.8ex] 
\end{tabular}}
\label{tab:exLagsSym}
\end{table}
\section{Restriction on FV due to Weak Gauge Symmetry}
\label{sec:genTheorems}
We have looked over various examples of FV, 
Hamiltonians, Lagrangians and the weak symmetries in Sec. \ref{sec:Lags}. 
Some Lagrangians meet the ``semiclassical'' weak 
symmetry, and some do not. Do the ``full quantum'' states realize the symmetry 
that the former Lagrangians possess? 
We know from Sec. \ref{sec:ex-simpIe-llusts} that the answer is not affirmative. 
So there may be a kind of restriction on quantum states.
This falls within the problem that Stone took up more than a decade ago.
\cite{Stone}
It is the problem of realizable FV in CS. 
\par
In this section we treat the problem 
of the restriction on types of spin CS, i.e. on those of the FV, 
in the full quantum picture when a spin CS Lagrangian has the weak semiclassical gauge $\psi$-symmetry (\ref{eqn:transf-Lag}) in the light of the formalism in Secs. \ref{sec:spinCSPI} and \ref{sec:symLagFV}. 
This means that we impose such restrictions on 
FV in spin CS that reflect the weak semiclassical gauge symmetry (\ref{eqn:transf-Lag}). 
It provides one answer to the mystery on spin CSPI 
posed by Stone in Ref. \refcite{Stone}. 
Or what we will perform is to see the results in 
Ref. \refcite{Stone} from a different point of view; 
from the viewpoint of the gauge symmetry of 
the action in spin CSPI. 
First, the general theorem, Theorem \ref{th:typeFV}, 
is given and proved in Sec. \ref{sec:th1}. 
Second, in Sec. \ref{sec:explicit-G}, we investigate 
the generator of the symmetry transformation, 
which yields Theorem \ref{th:Gform}. 
And then we revisit Theorem \ref{th:typeFV} 
via Theorem \ref{th:Gform}. 
It may help us to understand what is going on concretely.
\subsection{General results}
\label{sec:th1}
We  consider a class of Lagrangians which we have formulated in Sec. \ref{sec:symLagFV} and illustrated in Sec. \ref{sec:Lags}: 
We treat a Lagrangian with the weak gauge $\psi$-symmetry in which $A_3 = 0$ as well as $H({\bf\Omega})$ is invariant under the transformation (\ref{eqn:transfRS3}). 
Then the whole Lagrangian changes at most a total derivative. 
Among the examples in the preceding Sec. 
\ref{sec:ex-Lags}, 
(i), (ii), (vi) and (vii) meet the condition; 
see the Table \ref{tab:exLagsSym}. 
In the cases the following theorem holds. 
And thus actually only the FV in (i) survives; 
FV in the form of (ii), (vi) or (vii) are ruled out. 
The results agree with those indicated by Stone
\cite{Stone} from the viewpoint of two types of isotropy subgroups 
associated with semiclassical and full quantum dynamics. 
For the cases (iii), (iv) and (v) the theorem will not tell anything. 
\setcounter{theorem}{0}
\begin{theorem}
If a Lagrangian associated with SU(2) CS 
has the weak gauge symmetry related to the $\psi$-variable, 
the fiducial vector belongs to $\ket{m}$ or to 
the orbit of $\ket{m}$ under the action of 
${\hat R}({\bf\Omega})$ and
 the Dirac condition holds for $A_0$.
\label{th:typeFV}
\end{theorem}
\par
Before going into the following proof, 
we briefly mention in which direction we will proceed. 
Doing so is adequate for the purpose. 
\par
Since we concentrate on spin degrees of freedom, 
spin CS and FV, our Lagrangian differs from 
Refs. \refcite{Bal-NP}--\refcite{Aitch}. 
Moreover, as we stated above, 
we treat a rather wider class of Lagrangians 
including those illustrated in (i), (ii), (vi) and (vii) 
in Table \ref{tab:exLagsSym}. 
We look the problem quite generally 
in the light of CS and CSPI and our view includes the cases in Refs. \refcite{Bal-NP}--\refcite{Aitch} as special ones. 
The method by Aitchison in Ref. \refcite{Aitch}, however, applies also to ours. 
We extend it to our CS and CSPI with a rather wider class of Lagrangians. 
And thus we mainly proceed along Ref. \refcite{Aitch} 
in the following {\em with suitable changes}; 
we particularly observe how a state vector 
$\ket{\bf\Omega}$ behaves. 
That is our strategy. 
Let us start the proof now. 
\begin{proof}
Since we assume that the gauge $\psi$-symmetry 
holds for a Lagrangian, the Lagrangian takes the form of 
 (\ref{eqn:weak-sym-Lag}). 
And under the transformation (\ref{eqn:transfRS3}) 
a ket vector $\ket{\bf\Omega}$ and the Lagrangian 
change in the manners of (\ref{eqn:transf-ketV}) and 
(\ref{eqn:transf-Lag}), respectively. 
\par
Now, denote $\hat G$ such an operator 
that makes a transformation (\ref{eqn:transf-ketV}) 
on $\ket{\bf\Omega}$: 
\begin{equation}
\exp(- i {\hat G} \psi') \ket{\bf\Omega}
= \exp(- i {\hat G} \psi') {\hat R}({\bf\Omega}) \ket{\Psi_0}
= {\hat R}({\bf\Omega}) \exp(- i {\hat S}_3 \psi') \ket{\Psi_0}.
\label{eqn:transfG-ket}
\end{equation}
This means that $\hat G$ is the generator of 
(\ref{eqn:transf-ketV}) and also of (\ref{eqn:transf-Lag}) 
since we have assumed the FV and $\hat H$ meet the conditions on 
weak symmetry. 
In terms of $\hat G$, the change of $L$ 
becomes:
\begin{equation}
L \rightarrow L + \hbar \, 
\bra{\bf\Omega} {\hat G} \ket{\bf\Omega}
\, {\dot \psi'},
\label{eqn:transfG-Lag}
\end{equation}
where we have used the expression of $L$ in (\ref{eqn:action-pol}). 
Then we obtain from (\ref{eqn:transf-Lag}) and (\ref{eqn:transfG-Lag})
\begin{equation}
\bra{\bf\Omega} {\hat G} \ket{\bf\Omega}
 = A_0,
\label{eqn:GA_0}
\end{equation}
which is considered as a condition on semiclassical 
symmetry.
\par
From (\ref{eqn:transfG-ket}) we have the following relation 
between operators: 
\begin{equation}
\exp(- i {\hat G} \psi') {\hat R}({\bf\Omega}) 
= {\hat R}({\bf\Omega}) \exp(- i {\hat S}_3 \psi'). 
\label{eqn:transf-GRS3}
\end{equation}
For an infinitesimal transformation
we have 
\begin{equation}
{\hat G} {\hat R}({\bf\Omega}) 
= {\hat R}({\bf\Omega}) {\hat S}_3. 
\label{eqn:transf-inf-GRS3}
\end{equation}
The explicit form of 
$\hat G$, which we do not need in the present context, 
is given in Sec.
 \ref{sec:explicit-G}. 
Notice that (\ref{eqn:transf-inf-GRS3}) is {\em not identical with} 
the corresponding expression in Ref. 
\refcite{Aitch}. 
The latter is $[{\hat G}, {\hat R}({\bf\Omega}) ]
= {\hat R}({\bf\Omega}) {\hat S}_3$. 
\par
Let us consider a {\em full} 
quantum state so that the dynamics is conformable to semiclassical 
$\psi$-symmetry. 
In order to realize the state 
it is appropriate to borrow a standard field theoretic 
method \cite{Jackiw,Dirac-sub-cond}
 to our CS context. 
And thus we may perform the procedure by 
imposing an auxiliary condition on a state vector. 
The definition of $\hat G$ in (\ref{eqn:transfG-ket}), 
the consequent weak symmetry (\ref{eqn:transf-Lag}) 
and Noether's theorem tell us that the state vector is subject to a
 certain restriction in the sense of Dirac,
\cite{Dirac-sub-cond} 
and thus, following Ref. \refcite{Aitch}, we require a sort of Gauss' law:\cite{Jackiw} 
\begin{equation}
{\hat G} \, \ket{\bf\Omega} 
= A_0 \, \ket{\bf\Omega}.
\label{eqn:Gauss}
\end{equation}
Note that (\ref{eqn:Gauss}) is different from (\ref{eqn:GA_0}). It is clear that (\ref{eqn:Gauss}) is more stringent than 
(\ref{eqn:GA_0}). A set of $\set{\ket{\bf\Omega}}$ that satisfies (\ref{eqn:Gauss}) also meets (\ref{eqn:GA_0}). 
However, the reverse does not always hold. We will return to the relation between (\ref{eqn:GA_0}) and 
(\ref{eqn:Gauss}) in Sec. \ref{sec:anotherViewSCevolvs}. 
\par
As the finite form of (\ref{eqn:Gauss}) we have
\begin{equation}
\exp(- i {\hat G} \psi') \ket{\bf\Omega} 
= \exp(- i A_0 \psi') \ket{\bf\Omega}.
\label{eqn:weak-inv-vec1}
\end{equation}
On the other hand, with the aid of (\ref{eqn:transf-GRS3}), 
\begin{equation}
\exp(- i {\hat G} \psi') \ket{\bf\Omega} 
= {\hat R}({\bf\Omega}) \exp(- i {\hat S}_3 \psi') \ket{\Psi_0} 
= \sum_{m = - s}^{s} c_m \exp(- i m \psi') 
\ket{{\bf\Omega}, m}. 
\label{eqn:weak-inv-vec2}
\end{equation}
Combining (\ref{eqn:weak-inv-vec1}) with 
(\ref{eqn:weak-inv-vec2}) we obtain
\begin{equation}
\exp(- i A_0 \psi') \sum_{m = - s}^{s} c_m \ket{{\bf\Omega}, m} 
= \sum_{m = - s}^{s} c_m \exp(- i m \psi') 
\ket{{\bf\Omega}, m}. 
\label{eqn:weak-inv-vec3}
\end{equation}
Only a few FV meet the condition 
(\ref{eqn:weak-inv-vec3}), 
and indeed it is possible if and only if 
\begin{equation}
\ket{\Psi_0} = \ket{m}.
\label{eqn:FVm}
\end{equation}
The FV has only one nonzero component: 
The state vector becomes
\begin{equation}
\ket{\bf\Omega} = \ket{{\bf\Omega}, m}, 
\label{eqn:OOm}
\end{equation}
which is what we call ``rotated spin number states'' in {\bf I}.
From this we see 
\begin{equation}
A_0 = m. 
\label{eqn:A_0-m}
\end{equation}
Since $m = \cdots, - \frac12, 0, \frac12, \cdots$, 
Eq. (\ref{eqn:A_0-m}) indicates that the Dirac condition holds for $A_0$. 
\par
Let us put one more point about a FV. 
Decompose ${\hat R}({\bf\Omega}) 
= {\hat R}({\bf\Omega}') 
{\hat R}({\bf\Omega}'')
$
and we will obtain
\begin{equation}
\ket{\bf\Omega} 
= \ket{{\bf\Omega}'}
= {\hat R}({\bf\Omega}') 
\ket{\Psi_0 '}
\quad 
{\rm with}
\quad
\ket{\Psi_0 '} 
\equiv {\hat R}({\bf\Omega}'') \ket{m}. 
\label{eqn:FVRm}
\end{equation}
From this viewpoint 
we may interpret $\ket{\Psi_0 '}$ as a FV. 
Hence, from (\ref{eqn:FVm}) and (\ref{eqn:FVRm}), 
we see that the FV coincides with $\ket{m}$, one of the eigenstates 
of ${\hat S}_3$, or rides on the orbit of $\ket{m}$ 
under the SU(2) rotations. 
\end{proof}
The above Theorem \ref{th:typeFV} implies the strange 
feature of CSPI and FV: i.e., the Lagrangians, which depend upon the kinds of FV as shown 
in Table \ref{tab:exLagsSym}, in turn restrict them. 
The restriction condition determines the forms of FV. 
As a result, the FV belongs to 
$\ket{\Psi_0} = \ket{m}\ (m = s, s - 1, \cdots , -s)$, 
or can be reached from $\ket{m}$ by ${\hat R}^{(s)}({\bf\Omega})$. 
Mathematically, they are on orbits of $\ket{m}$ 
under the action of the SU(2) group.
\cite{Gil} 
And the Dirac condition
is permitted. 
This gives an answer to the problem posed in 
Ref. \refcite{Stone} in the light of 
our spin CSPI formalism. 
\par
Notice that the reverse statement does not hold: 
The Dirac condition does not always imply $\ket{\Psi_0} = \ket{m}$. 
This is apparent since a FV $\ket{\Psi_0} 
= \bigl((\frac23)^{1 /2}, 0, 0, (\frac13)^{1 / 2} \bigr)^T$ in (vi) 
and (vii) in 
Table \ref{tab:exLagsSym} yields $A_0 = \frac12$. 
\par
However, there is clearly an exceptional case: 
i.e. $A_0 = 0$ case in which 
the FV may have several nonzero components. 
\par 
Next, we revisit (\ref{eqn:weak-inv-vec3}) from another viewpoint. 
For a spin $s = 0, 1, 2, \cdots$ case, 
$m = 0, \pm 1, \pm 2, \cdots$. 
Putting $\psi' = 2 \pi$ in (\ref{eqn:weak-inv-vec3})
we have $\exp(2 \pi i A_0) = 1$, thus yielding 
$A_0 = n$ ($n$: integer). It is consistent with 
(\ref{eqn:A_0-m}). 
For a spin half-integer case, 
putting $\psi' = 4 \pi$ in (\ref{eqn:weak-inv-vec3}), 
we have $\exp(4 \pi i A_0) = 1$, 
and thus we obtain $A_0 = \frac12 n$ ($n$: integer). 
Since $m = \pm \frac12, \pm \frac32, \cdots$ 
in this case, 
the result agrees with (\ref{eqn:A_0-m}). 
We have already made a similar argument about 
semiclassical and full quantum propagators after 
(\ref{eqn:diff2props-FVm}).
Note that for a spin $\frac12$ case, 
it is always possible to write 
any FV in the form of $\ket{\Psi_0} = {\hat R}({\bf\Omega}_0) 
\ket{- \frac12}$ 
using a suitable Euler angle ${\bf\Omega}_0$: For spin $\frac12$ 
any FV can be reached from $\ket{- \frac12}$ 
 or $\ket{\frac12}$. 
We are able to describe any two-state system 
in terms of SU(2) CS. 
So the condition (\ref{eqn:FVRm}) holds automatically. 
What we have seen here is a CS version of 
magnetic charge quantization which has been known 
for a long time.
\cite{Aitch} 
\par
Now, it is widely known that the Dirac condition is 
related to the Dirac strings. 
The Dirac string, extending from the origin to half infinity 
in the $(\phi, \theta)$-space, corresponds to choosing 
the $\psi = \phi$ or $\psi = - \phi$ ``section'' 
\cite{WuYang} 
in the topological term (\ref{eqn:topTermSym}). 
In the present case it is surely possible to prepare the above $\psi$ since the gauge $\psi$-transformation promises the freedom. 
And the freedom, as we saw, comes from selecting the special type of FV: $\ket{\Psi_0} = \ket{m}$. 
Hence, {\em looking from the present spin CSPI formalism, whether the Dirac strings are permissible 
or not depends upon the types of FV.} 
\par 
One may observe that conventional arguments about the Dirac condition 
often imply that the particle interacting with a pole has spin $0$ or $\frac 12$, 
which falls within the type of FV that meets Theorem \ref{th:typeFV}. 
For a generic spin $s$, however, we have wider possibilities on FV. 
Then, if a Lagrangian made out of a FV does not possesses the gauge $\psi$-symmetry, we may expect 
wider possibilities also on magnetic charge. 
We will stress it again in Sec. \ref{sec:discussion}. 
It is, of course, an open question whether 
we can apply our discussion to real magnetic monopoles. 
However, the approach presented here may provide us with a fine view 
of real monopoles since real and fictitious monopoles enjoy common 
mathematical descriptions. 
\subsection{Explicit form of the generator $\hat G$}
\label{sec:explicit-G}
In the preceding subsection we used the operator $\hat G$ 
satisfying (\ref{eqn:transfG-ket}) or (\ref{eqn:transf-GRS3}). 
Although we have not needed its explicit form, a natural question arises: What does it look like? 
We find the explicit form of the generator $\hat G$ easily in Theorem \ref{th:Gform} below. 
This may be a by-product of Theorem \ref{th:typeFV}. 
Conversely, however, the form brings us to Theorem \ref{th:typeFV} via
another route again. 
Hence Theorem \ref{th:Gform} helps us to understand 
Theorem \ref{th:typeFV} much deeper. 
%
\begin{theorem}
For an operator $\hat G$ to satisfy 
{\em  (\ref{eqn:transfG-ket})} or 
{\em  (\ref{eqn:transf-GRS3})} it is necessary 
and sufficient that $\hat G$ is expressed as 
\begin{equation}
{\hat G} 
= {\hat R} ({\bf\Omega})
{\hat S}_3 
{\hat R}^{+} ({\bf\Omega})
= {\bfm S}{\bf \cdot} {\bfm n}, 
\label{eqn:G}
\end{equation}
where ${\bfm n} \equiv (\sin\theta \cos\phi, 
\sin\theta \sin\phi, \cos\phi)$. 
\label{th:Gform}
\end{theorem}
\noindent
\begin{proof}
Let us see the necessity first. 
From (\ref{eqn:transf-GRS3}) we have 
\begin{equation}
\exp(- i {\hat G} \psi') 
= {\hat R}({\bf\Omega}) \exp(- i {\hat S}_3 \psi') {\hat R}^{+}({\bf\Omega}) 
= \exp[ - i {\hat R}({\bf\Omega}) {\hat S}_3 {\hat R}^{+}({\bf\Omega}) \psi' ], 
\label{eqn:transf-GRS3-A}
\end{equation}
which, with the help of 
({\bf I}-A.7), results in (\ref{eqn:G}). 
Next, that (\ref{eqn:G}) is sufficient is obvious since 
we only need to cross the last equality in 
(\ref{eqn:transf-GRS3-A}) in reverse order.
\end{proof}
We see that the generator $\hat G$ is nothing but a sort of generalized ``Hopf map''. 
The usual Hopf map is given by 
${\hat R} ({\bf\Omega})
{\hat \sigma}_3 
{\hat R}^{+} ({\bf\Omega})
= {\bfm \sigma} {\bf \cdot} {\bfm n}$.
\cite{Aitch} 
It is often referred to as the quantity 
which indicates the gauge $\psi$-symmetry, 
and it is also exactly what is taken 
 as the tool of ``gauge fixing'' in 
Refs. \refcite{Bal-NP} and \refcite{Bal-CTQS}. 
\par
Next we revisit Theorem \ref{th:typeFV} 
via Theorem \ref{th:Gform}. 
Now, ${\hat S}_3$ is a $(2 s + 1) \times (2 s + 1)$
 matrix whose eigenvectors are completely given by 
\begin{equation}
{\hat S}_3 \ket{m} = m \ket{m}.
\label{eqn:S3-EV}
\end{equation}
Then, operating a nonsingular matrix ${\hat R}({\bf\Omega})$ 
on both sides of (\ref{eqn:S3-EV}), we have, 
employing (\ref{eqn:G}),
\begin{equation}
{\hat G} \, \ket{{\bf\Omega}, m} = m \ket{{\bf\Omega}, m},
\label{eqn:GOm}
\end{equation}
which specifies all the $(2 s + 1)$ eigenvectors of 
$\hat G$ thoroughly. 
Comparing (\ref{eqn:GOm}) with (\ref{eqn:Gauss}), 
we realize that $\ket{\bf\Omega}$ must coincide with one of the
$\set{\ket{{\bf\Omega}, m}}$. 
This leads us again to (\ref{eqn:OOm}), 
from which we consequently obtain 
(\ref{eqn:FVm}) and (\ref{eqn:A_0-m}). 
We thus confirm that (\ref{eqn:FVm}) and 
(\ref{eqn:A_0-m}) hold again. 
\par 
Next, we put two additional comments. 
First, the form of (\ref{eqn:G}) can be obtained 
also by the infinitesimal 
relation (\ref{eqn:transf-inf-GRS3}). 
Since (\ref{eqn:transf-inf-GRS3}) is independent of $s$, 
we may try $2 \times 2$ matrices:
\begin{equation}
{\hat G}
\equiv \alpha_{+} {\hat S}_{+} + \alpha_{-} {\hat S}_{-} 
+ \alpha_{3} {\hat S}_{3} 
= \left(
\begin{array}{lc}
\frac12 \alpha_3 
& \alpha_+
\\
\alpha_-
& - \frac12 \alpha_3
\end{array}
\right)
\label{eqn:G-1/2}
\end{equation}
and ${\hat R}^{(1 / 2)} ({\bf\Omega})$; 
see ({\bf I}-A.1) for the expression of ${\hat R}^{(1 / 2)} ({\bf\Omega})$. 
In this manner we obtain (\ref{eqn:G}) again. 
It is clear that the direct evaluation of 
$\exp( - i \psi' {\hat G})$ using 
(\ref{eqn:G-1/2}) leads to (\ref{eqn:transf-GRS3}) 
as well; see Refs. \refcite{Gil} and \refcite{ACGT} for such matrix calculations.
Second, we point out that $A_0$ in (\ref{eqn:defA}), 
(\ref{eqn:GA_0}) and (\ref{eqn:GOm}) are mutually consistent. 
\section{Another View of Semiclassical Motions}
\label{sec:anotherViewSCevolvs}
In Ref. \refcite{Stone} Stone characterized 
CS, FV and the consistency between semiclassical and 
full quantum dynamics by 
a slightly different way from ours. 
In order to see what they looked like, 
he introduced the isotropy subgroup ${\cal H}_0$ 
for a given Lie group $\cal G$ that stabilized 
the expectation values of the Lie group generators in the state of a FV 
in addition to the usual isotropy subgroup $\cal H$ 
in (\ref{isotropySG}). 
He showed that semiclassical orbits lived on ${\cal G} / {\cal H}_0$, whereas the full quantum dynamics was governed by ${\cal G} / {\cal H}$.
\par 
Let us concentrate on ${\cal H}_0$ here. 
For the present spin CS case it reads: 
\begin{equation}
{\cal H}_0 
= \{ h \in {\cal G} | \, 
\bra{h} {\hat S}_i \ket{h}
 = \bra{\Psi_0} {\hat S}_i \ket{\Psi_0}
\},
\qquad \bigl( \, \ket{h} \equiv h \ket{\Psi_0}, \, i = \pm, 3\bigr). 
\label{eqn:isoFV2SC}
\end{equation} 
Notice that $\ket{0}$ is used in place of $\ket{\Psi_0}$
and ${\cal H}_{\ket{0}}$ instead of $\cal H$ in 
Ref. \refcite{Stone}. 
We take latter notations to
keep harmony with the expressions in the preceding sections. 
One may feel that the relation between the description by ${\cal H}_0$ and that by $\cal H$ resembles 
the connection between (\ref{eqn:GA_0}) and 
(\ref{eqn:Gauss}). 
One describes the semiclassical dynamics. 
And the other is related to full quantum time evolution. 
\par
We now revisit the framework of Ref. \refcite{Stone} 
from our point of view in Sec. \ref{sec:spinCSPI}--Sec. \ref{sec:genTheorems}. 
Let us take up three representative types of FV 
which are numbered (i), (ii), (iv) in Table \ref{tab:exLagsSym}. 
Then we obtain the following results due to Ref. 
\refcite{Stone} 
illustrated in Table \ref{tab:isotropySG-FV}:
\begin{table}[htbp] 
\tbl{Two types of isotropy subgroups and FV}
{\begin{tabular}{@{}cccc@{}}
 \Hline 
\\[-1.8ex] 
 & (i) & (ii) & (iv) \\[0.8ex] 
\hline\\[-1.8ex] 
\\
 FV & $\ket{m} $ 
& $\begin{pmatrix} \sqrt{2 / 3} \\ 0 \\ 
\sqrt{1 / 3} 
\end{pmatrix}$ 
 & $\begin{pmatrix} \sqrt{1 / 3} \\ 
\sqrt{1 / 3} \\ 
\sqrt{1 / 3} \end{pmatrix}$ \\
\\
 $\cal H$ & $\set{ \exp(- i \psi' {\hat S}_3)}$
 & $\set{\bf 1}$ & $\set{\bf 1}$ \\
\\
 ${\cal H}_0$ & $\set{ \exp(- i \psi' {\hat S}_3)}$ & $\set{ \exp(- i \psi' {\hat S}_3)}$ 
& $\set{\bf 1}$ \\
\\
\Hline\\[-1.8ex] 
\end{tabular}}
\label{tab:isotropySG-FV}
\end{table}

%
\noindent
In cases (i) and (ii) we have 
${\cal H}_0 = \set{h} 
= \set{ \exp(- i \psi' {\hat S}_3)}$. 
It is clear that the effect of ${\cal H}_0$ in the Lagrangians amounts to 
that of (\ref{eqn:transfRS3}) which 
features the weak $\psi$-symmetry. 
In fact we have used the convention and notation ``$- \psi'$'' 
so that one can see the accordance easily. 
Contrary, we have ${\cal H}_0 = \set{\bf 1}$ in the case
 (iv) and the case has nothing to do with the weak $\psi$-symmetry. 
Hence we are going to investigate (i) and (ii) with the aid
of (\ref{eqn:isoFV2SC}) instead of (\ref{eqn:transfRS3}). 
Consider the topological term first. 
For a FV with parameter dependence we have 
a more appropriate interpretation of ({\bf I}-35) as:
\begin{eqnarray}
{\hat R}^{+}({\bf\Omega}) 
\frac{\partial}{\partial t} 
{\hat R}({\bf\Omega}) 
& = & {\hat R}^{+}({\bf\Omega})
 \left( \frac{\partial} {\partial t} 
 {\hat R}({\bf\Omega}) \right) 
+ \frac{\partial} {\partial t}
\nonumber \\
& = & - i ({\dot \phi} \cos\theta + {\dot \psi}){\hat S_3}
+ \frac12 ( i {\dot \phi} \sin\theta - {\dot \theta} ) 
\exp(i \psi){\hat S_+} 
\nonumber \\
& & \qquad 
+ \frac12 ( i {\dot \phi} \sin\theta + {\dot \theta} ) 
\exp(- i \psi){\hat S_-} + \frac{\partial} {\partial t}. 
\label{eqn:pol-R2}
\end{eqnarray}
Since (\ref{eqn:pol-R2}) is linear in ${\hat S}_i$, 
we observe that under the operation ${\cal H}_0$ on $\ket{\Psi_0}$ the
following relation holds: 
\begin{eqnarray}
\bra{\Psi_0} {\hat R}^{+}({\bf\Omega}) 
( \partial / \partial t) 
{\hat R}({\bf\Omega}) 
\ket{\Psi_0} 
& \rightarrow & 
\bra{h} {\hat R}^{+}({\bf\Omega}) 
( \partial / \partial t) 
{\hat R}({\bf\Omega}) 
\ket{h} 
\nonumber \\
& = & \bra{\Psi_0} {\hat R}^{+}({\bf\Omega}) 
( \partial / \partial t) 
{\hat R}({\bf\Omega}) 
\ket{\Psi_0} 
- i {\dot \psi'}, 
\label{eqn:transfTopolTermH_0}
\end{eqnarray} 
where we have used (\ref{eqn:isoFV2SC}). 
Hence, the topological term $\bra{\bf\Omega} 
(\partial / \partial t) \ket{\bf\Omega}$ 
is weakly symmetric under ${\cal H}_0$. 
The result may be obvious, for it is clear that (\ref{eqn:isoFV2SC}) corresponds to (\ref{eqn:transfRS3}).
Besides, assuming ${\hat H} = {\hat H}_{\rm NMR}$ in (i), $\hat H$ is also linear in 
${\hat S}_i$ in the present cases (i) and (ii) in Tables 
\ref{tab:exLagsSym} and \ref{tab:isotropySG-FV}, 
and thus $\bra{\bf\Omega} {\hat H} \ket{\bf\Omega}$ is invariant under 
${\cal H}_0$. 
Then the whole Lagrangians possesses the weak $\psi$-symmetry. 
So we have confirmed that the condition (\ref{eqn:isoFV2SC}) reproduces 
one about the weak gauge $\psi$-symmetry for a
Lagrangian discussed in Sec. \ref{sec:symLagFV}. 
\par
If one tries to express the complete symmetry 
of Hamiltonians that have such higher 
order products of ${\hat S}_i$ as ${\hat H}_{\rm NQR}$ in (\ref{eqn:H-NQR}), more stringent conditions like 
\begin{equation}
{\cal H}_0 
= \{ h \in {\cal G} | \, 
\bra{h} 
{\hat S}_{i_1} \cdots {\hat S}_{i_\ell}
 \ket{h}
 = \bra{\Psi_0} 
{\hat S}_{i_1} \cdots {\hat S}_{i_\ell}
\ket{\Psi_0}
\}
\qquad (i_1, \cdots, i_\ell = \pm, 3)
\label{eqn:isoFV2SC2}
\end{equation}
may be required. 
Here ${\hat S}_{i_1} \cdots {\hat S}_{i_\ell}$ 
is an arbitrary $\ell$ product of 
${\hat S}_\pm$ or ${\hat S}_3$ that appears in the 
Hamiltonian. For example, we know ${\hat H}_{\rm NQR}$ gives $\ell =2$ 
and (\ref{eqn:isoFV2SC2}) contains the condition 
$
\bra{h} ({\hat S}_{+})^2 \ket{h}
 = \bra{\Psi_0} ({\hat S}_{+})^2 \ket{\Psi_0}
$. 
Augmented with (\ref {eqn:isoFV2SC2}) 
we see that the description due to ${\cal H}_0$ 
brings a criteria for the weak gauge symmetry 
which works on all the cases in Table \ref{tab:exLagsSym}. 
\par
We have thus looked over two types of 
descriptions on CS, FV and semiclassical evolutions: 
One uses the shift of the $\psi$ variable in 
${\hat R}({\bf\Omega})$ as (\ref{eqn:transfRS3}) and
the other, i.e., (\ref{eqn:isoFV2SC}) and 
(\ref{eqn:isoFV2SC2}), employs a transformed FV. 
They equally describe semiclassical dynamics well. 
And no matter what description we choose, 
both Ref. \refcite{Stone} and Theorem \ref{th:typeFV} in 
Sec. \ref{sec:genTheorems} tell us 
that if we have the gauge $\psi$-symmetry, we are led to 
a standard FV and informative CS in the full quantum dynamics.
\section{Discussion}
\label{sec:discussion}
We have studied Lagrangians having a weak gauge symmetry 
in the light of spin CSPI with a general FV. 
We have set a condition on a state vector 
in order that the full quantum description keeps the semiclassical symmetry. 
This gives the restriction on FV. 
Then the types of CS get limited. 
It is mandatory that the spin CS ride on 
the orbits of $\ket{m}$ under the action of 
${\hat R}({\bf\Omega})$ in the full quantum dynamics. 
And the fictitious monopole charge $A_0$ is so quantized 
as to bring the Dirac condition. 
Otherwise, the $\psi$-variable becomes ``anomalous''. 
Of course in natural sciences it is Nature who gives a final decision. 
However, we expect that the results are all right in the case. 
Concerning the matter, we find that the results agree with those due to Stone,\cite{Stone} 
who first posed the problem in a general 
framework of CS and FV. 
\par
Notice that {\em the rule to determine $A_0$, 
i.e. the monopole charge or strength, is not built in the CSPI a priori, 
but we impose it from physical demands 
--- the gauge $\psi$-symmetry.} 
This is the way in which spin CSPI with general FV 
bring the Dirac condition for fictitious monopole charges. 
Remember that the monopole charge quantization 
condition is not derived by the Schr{\"o}dinger equation itself, 
but by boundary or topological conditions 
also in the usual wave function formalisms.
\cite{WuYang,Dirac,Dirac-IJTP} 
\par
Now, let us see a future prospect. 
First, in the present paper, we have treated Lagrangians 
with the gauge $\psi$-symmetry. 
If we consider a Lagrangian without the symmetry, 
the situation looks rather different. 
See, for example, (iii), (iv) and (v) in Table \ref{tab:exLagsSym}. 
In the cases, we could not impose a subsidiary condition, 
i.e. Gauss' law, which produces the quantization 
of a fictitious monopole charge $A_0$ anymore. 
Theorem \ref{th:typeFV} does not bring us any information for the case. 
And then what would happen to a FV, $\ket{\Psi_0}$, and $A_0$? 
The problem, as well as physical applications of the 
present case, seems to be so intriguing for future investigations. 
\par
Second, it is clear that the present formalism 
may be extended to wider CSPI cases. 
Among them SU(1, 1) and SU(3) cases 
sound most probable candidates. 
The latter case is, of course, related to QCD. 
What do the restrictions on FV give physical systems described in terms
of these CS? Do they bring new information 
other than the previous ones?
\cite{Bal-NP}${}^{\mbox{--}}$\cite{Aitch} 
In this respect it might be as well to remember that 
Stone \cite{Stone} actually discussed the restriction on FV 
for CS constructed from wider Lie groups. 
\par
Third, we give subsidiary comments on CS and FV: 
One might observe a close formal analogy between 
CS with general FV and the ground states of many body systems 
or vacua in field theory.
\cite{NamPol,Aitch-GFT,Ryder} 
At least mathematical apparatus, a Lie group $\cal G$ 
and its coset space ${\cal G} / {\cal H}$, 
are common to both of the cases.
\cite{Ryder} 
FV look like ground states or vacua. 
Of course, in the many body systems the ground states themselves 
are expressed in terms of CS. However, such ground states are also absorbed into arbitrary FV. 
Moreover, we may be able to prepare room for dealing symmetries associated with higher energy levels than vacua with the aid of the present arbitrary FV formalism. 
It is an open question as to whether there are some deeper implications behind the formal resemblance. 
In addition we want to indicate one more point on CS and FV. It is on the
definitions of two types of isotropy subgroups ${\cal H}_0$ and $\cal H$ in Sec. \ref{sec:anotherViewSCevolvs}. 
They correspond to the relation between (\ref{eqn:GA_0}) and (\ref{eqn:Gauss}). 
It seems natural to feel that they remind us of subsidiary conditions 
in covariant quantization of photons due to 
Gupta--Bleuler.\cite{MandlShaw} 
In that case the expectation value of the operator 
describing the Lorentz condition for a state vector is more crucial than the effect of the operation on the vector itself 
to establish the connection between quantized photon field and 
classical electrodynamics. 
And a certain gauge transformation leaves the expectation value
invariant. 
This is also the case for the above-mentioned vacua in field theory. 
There again the expectation value of a field operator plays a central role. 
We have not yet known whether there exists any deeper meaning of the analogies or not either.
\par
Fourth, let us look into a slightly different direction.
\footnote{The contents of the paragraph grew out of a Reviewer's suggestion.} 
The subject is on the connection between the results here and QC or QI.
\cite{NielChuang}  
Our results have been concerned with CSPI and gauge symmetry, and thus it seems that they have nothing to do with QC and QI directly.
Of course, as we pointed out in {\bf I}, there is a possibility that the CS with arbitrary FV may appear in QC. 
However, we now intend to proceed to another point. What we want to indicate is that changing a FV 
somewhat resembles choosing a reference frame in relativity. 
In this respect it should be noted that the nature of entanglements in QI changes according to choosing reference 
frames, which are inertial or noninertial.
\cite{GinAdami,AlsMilburn} 
Making a unitary rotation on a FV means that we choose a different base like in  (\ref{eqn:FVRm}). 
Then, we see in general the weights of 
the $A_0$ and $A_3$ terms also change. 
Although the entanglements in $A_3$-term  are those of coefficients of a FV, the situation is somewhat similar 
to that in entanglements for
 inertial frames.
\cite{GinAdami} 
Next, a nonstandard FV that is not reached by 
${\hat R}(\bf\Omega)$ from $\ket{m}$ may correspond to choosing a noninertial frame. 
In both cases there may be a connection between our 
$A_0$, $A_3$-terms and teleportation concurrence and 
fidelity in QI.
\cite{GinAdami,AlsMilburn} 
Besides $\bra{h}{\hat S}_3 \ket{h}$ in (\ref{eqn:isoFV2SC}) looks like a quantity that plays the role for indicating the teleportation fidelity. In relativity cases, Lorentz 
boosts and  Bogoliubov transformations appear in inertial frames and 
noninertial ones, respectively. 
We know that both transformations are related to the SU(1, 1) group. 
Therefore, extending the methods in {\bf I} 
to SU(1, 1) CS based on a general FV 
proposes a new specialized clue to those problems. 
And our symmetry viewpoint may shed a new light 
in understanding fidelity in QI. 
Especially one for the inertial frame.
\par
Finally, we know that condensed matter systems 
have possibilities to simulate monopoles and gauge field theories.
\cite{Aitch-GFT,MSW}${}^{\mbox{--}}$\cite{QLZ2} 
Monopole-type fields may appear in molecular physics,
\cite{MSW} superfluid helium,\cite{SalVol,VollWolfle} 
and a topological insulator.
\cite{QLZ2}
Condensed matter physics 
makes it possible to examine such concepts 
in laboratories. 
The problems of CS and gauge symmetry discussed here 
will be of interest for a variety of realms of physics 
including condensed matter physics. 
For symmetry is one of the basic principles that 
penetrate all of physics.
\section*{Acknowledgments}
The author is grateful to Prof. M. Stone for 
sending the draft of Ref. \refcite{Stone} 
more than a decade ago, 
which has drawn his attention to the problem 
discussed in the present article. 
\section*{References}

 \end{document}